\documentstyle[12pt, epsfig]{article}

\def\Re{\,\hbox{Re}\,}

\def\epm#1#2{\hbox{${\lower1pt\hbox{$\scriptstyle +#1$}}
\atop {\raise1pt\hbox{$\scriptstyle -#2$}}$}}

\def\gsim{\mathrel{\rlap{\lower4pt\hbox{\hskip1pt$\sim$}}
    \raise1pt\hbox{$>$}}}         

\def\frac#1#2{{{#1}\over {#2}}}

\def\as{\alpha_s}

\def\MS{\hbox{$\overline{\rm MS}$}}

\catcode`@=11 
\renewcommand\section{\@startsection {section}{1}{\z@}
    {-3.5ex plus -1ex minus -.2ex}{2.3ex plus .2ex}{\bf}}
\renewcommand\subsection{\@startsection {subsection}{1}{\z@}
    {-3.5ex plus -1ex minus -.2ex}{2.3ex plus .2ex}{\it}}
\def\slash#1{\mathord{\mathpalette\c@ncel#1}}
 \def\c@ncel#1#2{\ooalign{$\hfil#1\mkern1mu/\hfil$\crcr$#1#2$}}
\def\lsim{\mathrel{\mathpalette\@versim<}}
\def\gsim{\mathrel{\mathpalette\@versim>}}
 \def\@versim#1#2{\lower0.2ex\vbox{\baselineskip\z@skip\lineskip\z@skip
       \lineskiplimit\z@\ialign{$\m@th#1\hfil##$\crcr#2\crcr\sim\crcr}}}
\catcode`@=12 

\def\be{\begin{equation}}
\def\ee{\end{equation}}
\def\bea{\begin{eqnarray}}
\def\eea{\end{eqnarray}}

\begin{document}
\begin{titlepage}
\setcounter{page}{0}
\begin{flushright}
{\tt hep-ph/9812382}\\
{RM3-TH/99-3}\\
\end{flushright}
\begin{center}
{\Large \bf Recent developments in deep-inelastic scattering}\\

\vskip 1.5cm

{\bf Stefano Forte}\footnote{On leave from INFN, Sezione
di Torino, Italy}\\
{\sl INFN, Sezione di Roma III}\\
{\sl Via della Vasca Navale 84, I-00146, Roma, Italy}\\
\end{center}
\vskip 2.cm
\begin{abstract}

We review several recent 
developments in the theory and phenomenology
of deep-inelastic scattering, with particular emphasis on precision
tests of QCD and progress in the detailed perturbative 
treatment of structure
functions and parton distributions. We discuss specifically
determinations of $\alpha_s$;
higher twist contributions to structure functions and renormalons;
structure functions at small $x$ and resummation of energy logarithms.
\end{abstract}
\vfill
\leftline{July  1999}
\end{titlepage}
\section{DIS in the collider age}
Deep-Inelastic Scattering (DIS) is the traditional testing ground of
perturbative QCD. In the last several years, perturbative QCD has 
become, as an integral part of the standard model, a firmly
established theory, which allows performing detailed and reliable
calculations. Current work is focussed on
the precise determinations of the unknown parameters of the theory,
and  the development of reliable computational techniques. The
parameters to be determined include not only the
 strong 
coupling $\alpha_s$, which is the only free parameter in the
QCD Lagrangian, but also all quantities
which are determined from the nonperturbative low--energy
dynamics, and thus, even
though  in principle computable, must in practice be treated as
a phenomenological input in the perturbative domain.

In the case of DIS, the low--energy parameters
are the parton distributions 
of hadrons.
Thanks to the operator-product expansion and the
renormalization group, or more in general factorization theorems, the
coefficient functions which relate parton distributions to
measurable cross-sections can be computed in perturbation theory. Furthermore, 
the scaling violations of the parton distributions are 
related to the short--distance behavior of local operators in the theory, and
can thus also be computed reliably.\footnote{The theory of DIS is a
textbook\cite{books} subject. Here we will follow the notation and conventions
of Ref.\cite{guido}.}
Deep-inelastic scattering provides 
therefore the most accurate way of determining the parton distributions which 
are then used as an input in the perturbative computation of hard processes 
relevant for collider physics, as well as, through scaling violations, 
one of the prime ways of determining $\alpha_s$. 

In recent times, the status of the DIS phenomenology has changed substantially, 
largely due to the advent of 
the HERA lepton-hadron collider, which has enormously enlarged
the kinematic coverage of the data. In the simplest fully inclusive
case, 
the DIS kinematics is fully parametrized by the virtuality $Q^2$ of the
photon which mediates the lepton-hadron process, and the total
center-of-mass energy 
\be
W^2=Q^2{1-x\over x}
\label{kin}
\ee
of the virtual photon-hadron
process
(which  implicitly defines the Bjorken variable $x$). By
definition, the DIS region correspond to large  $Q^2$; the
accessible values of $x$ are then limited by the available
energy. HERA has thus opened up a large previously unaccessible
region in the $(x,Q^2)$ plane.
However, some 
very interesting complementary 
information has also been obtained recently by fixed-target experiments, which
have provided high-statistics data especially in the polarized case,
in neutrino scattering experiments, and in charged lepton scattering in 
the large $x$ region,
where $W^2$ 
is low on the scale of $Q^2$.
\begin{figure}
\centering\mbox{
    \epsfig{file=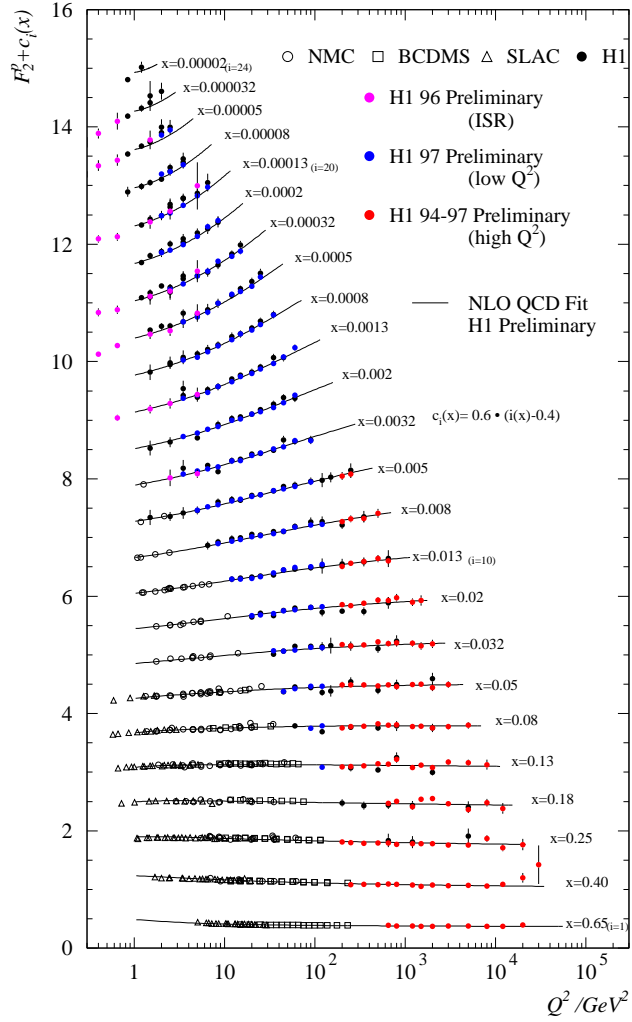,height=11cm}}
 \caption[]{The proton
structure function $F_2(x,Q^2)$ measured by the H1 collaboration at 
HERA as well as by SLAC\cite{slac}, BCDMS\cite{bcdms} and NMC\cite{nmc}
fixed-target experiments (from Ref.\cite{h1van}).}
\label{H1F2}
\end{figure}

The current state of the art  can be inferred
 from Fig.~1, which displays the 
data on the inclusive structure function $F_2(x,Q^2)$
collected by one of the HERA collaborations between 1994-1997, as well as  
earlier fixed-target data. The data are compared to the behavior 
expected by solving the Altarelli-Parisi evolution equations to 
next-to-leading order (NLO), with 
a fixed value of the strong coupling 
$\alpha_s(M_z)=0.118$. Parton distributions are parametrized at a
 fixed initial scale $Q_0^2=2$~GeV$^2$ with a dozen free parameters,
 fitted by comparing to the data.
It is apparent that despite the enormous magnification of kinematic
coverage, there is perfect agreement between the data and the QCD
prediction, and the agreement does not seem to deteriorate even  when
approaching regions where the NLO computation
might be expected to be not entirely
adequate. In particular, no signs of deviation from the  NLO QCD
prediction are seen in the small $x$ region, which was unexplored
before HERA.

It is important to understand what is
and what is not remarkable about the agreement of data and theory
displayed in Figure 1.
The fact that it is possible to achieve a good description of a large
number of data with relatively few parameters is not especially
interesting {\it per se}: it may simply indicate that the functional
dependence of the data on the kinematic variables is sufficiently
smooth. 

In fact,  
there exist parameterizations of $F_2(x,Q^2)$, constructed by
experimental collaborations for the purpose of data
analysis, which
interpolate all available data~\cite{tul}. 
These parameterizations do not use perturbative
QCD at all: they simply fit a functional form to the data, yet they 
hardly need a larger number of free parameters. Furthermore,
 several recent parameterizations of $F_2$ at low $Q^2$
(where perturbative evolution is not valid), derived within various
non-perturbative low energy models\cite{mods} 
 appear to give a reasonably good description of the $F_2$
data 
(though generally not quite as accurate as the QCD calculation) 
even up to values of $Q^2$ as large as 100~GeV$^2$, where perturbative
QCD should be used instead. 
Conversely, a recent phenomenological parameterization\cite{haidt},
inspired by a linearization of the perturbative QCD prediction, 
describes the low $x$ and large $Q^2$ data with only 4 parameters, and
turns out to work very well even when $Q^2$ is as low as 0.05~GeV$^2$,
where the perturbative approach is certainly invalid. 

It would
be equally unreasonable to take the latter as evidence of the validity
of perturbation theory at very low $Q^2$, or the former as
evidence that at large $Q^2$ perturbative QCD should be replaced by
nonperturbative
models: what distinguishes a perturbative computation  such as the one shown
in Figure~1 is its predictive power. Namely, the
 data are parametrized as a
function of $x$ at a reference scale $Q_0^2$, and increasing the number
of parameters only results in a more accurate description of the $x$
dependence of the data at that
scale. 
 The structure function at any other scale is then {\it predicted}, by
solving the evolution equations. This is to be contrasted with the
situation in a fit, whether or not inspired by a theoretical model.
There,  no prediction is available:
 the $x$ and $Q^2$ dependence are fitted
in a certain kinematic domain, and then extrapolated outside it.

The extraordinary success of the perturbative prediction displayed in Fig.~1
reflects the fact that the meaning of a ``test of QCD'' has changed in
the last decade. The correctness and consistency 
of the theory are no longer at stake: tests of its parameter-free
predictions (such as the Bjorken sum rule~\cite{abfr}, Sect 2.2, or
the small $x$ slope~\cite{test}, Sect.2.3) 
are impressively succesful. The phenomenological emphasis
is now on precision determination of parameters which must be used as
input for the reliable calculation of hadron collider processes, and
as backgrounds to possible new physics. As an example of this, we will
discuss in Sect.~2 the determination of $\alpha_s$, where the
agreement of values extracted from different experiments is now taken
for granted, and the aim of current work
is to obtain an accurate value of $\alpha_s$
for use in the computation of standard model processes. On the other
hand, ``testing'' the theory has taken the meaning of 
stretching the usual perturbative
techniques based on hard factorization. This can be done moving towards
regions of lower scale, where power corrections are important, as we
will discuss in Sect.~3 with specific reference to DIS at large $x$. 
It can also be done by studying regions where more than one hard scale
is present, as we will discuss in Sect.~4 in the case of small $x$
evolution.
An altogether different possibility is to stretch the domain of
factorization theorem to less inclusive phenomena. Specifically,
recent data at HERA (and elsewhere) 
have greatly extended our
knowledge of lepton-hadron scattering 
in the diffractive region, both within and outside
the deep-inelastic domain.
A discussion of this topic\cite{diff} 
is however outside the scope of this 
review.\footnote{Diffraction in the DIS region can be included in a
perturbative framework by introducing fracture
functions\cite{frac}, 
which allow a separation of the nonperturbative information
from the perturbative one\cite{deflor} 
and may prove a useful way to arrive at a
reliable theory of diffraction.}

\section{Determinations of $\alpha_s$}
A glance to the current edition of the particle data book\cite{pdb}
shows that a sizable part of our knowledge of $\alpha_s$
comes from DIS, and specifically from scaling violations. 
Even though this kind of  determination of $\alpha_s$ does not have nominally
the smallest uncertainty, it is, together with the determination from
the $Z$ width, the most reliable from a theoretical
viewpoint. The main recent progress here, in comparison for instance
to the previous edition\cite{pdbold} of the particle data book,
consists of the fact that, even though the experimental uncertainty
has been only marginally reduced, 
the central value of $\alpha_s(M_z)$ is 
now in significantly better agreement with the world average, 
and with
the value obtained from LEP data. This is partly due to the fact that
some pieces of data had  
experimental problems which have
now been corrected, partly due to the fact that some sources of
theoretical uncertainty hadn't been appreciated fully, and
partly to the availability of new data.
Here, we will review recent advances in the extraction of
$\alpha_s$ from inclusive structure function data. Of course,
$\alpha_s$ can
also be extracted from jet rates and event shapes. These
determinations, which are only possible at large center-of-mass
energy, in the lepton-hadron case have first  become possible at
HERA; they are however still in their infancy and not yet competitive
with inclusive determinations\cite{asex}

\subsection{Scaling violations of structure functions}
The scaling violations of the
structure functions $F_2$ and $F_3$
(unpolarized) and $g_1$ (polarized) are
driven by the value of $\alpha_s$ according to
\bea
{d\over dt} F_2 (N,Q^2)&=&
{\alpha_s(Q^2)\over
2\pi} \left[\gamma_{NS}(N) F_2^{\rm nonsing.}+
\gamma_{qq}(N) F_2^{\rm sing.}\right.\nonumber\\
&&\quad\left.\phantom{g_1^{\rm nonsing.}}+2 c \, 
n_f\gamma_{qg}(N)g(N,Q^2)\right]+O(\alpha_s^2)\nonumber\\
{d\over dt} F_3 (N,Q^2)&=&
 {\alpha_s(Q^2)\over
2\pi} \gamma_{NS}(N) F_3 +O(\alpha_s^2)
\\
{d\over dt} g_1 (N,Q^2)&=&
 {\alpha_s(Q^2)\over
2\pi} \left[\Delta \gamma_{NS}(N) g_1^{\rm nonsing.}+
\Delta \gamma_{qq}(N) g_1^{\rm sing.}\right.\nonumber\\
&&
\quad\left.\phantom{g_1^{\rm nonsing.}}
+c\, n_f\Delta \gamma_{qg}(N)\Delta g(N,Q^2)\right]+O(\alpha_s^2),
\label{scv}
\eea
where  $F_2(N, Q^2)\equiv \int_0^1\,dx x^{N-1} F_2(x,Q^2)$ 
is the $N$--th Mellin moment of the function $F_2(x,Q^2)$,
$g$ and $\Delta g$ are the unpolarized and polarized gluon
distributions, the
anomalous dimensions $\gamma(N)$ are the Mellin transforms
of the Altarelli-Parisi splitting functions $P(x)$,
and the factor $c={1\over n_f} 
\sum_{i=1}^{n_f} e^2_i$ for charged lepton and $c=1$ for neutrino scattering.
In order to extract $\alpha_s$ we need thus either to be able to
separate out the nonsinglet component of $F_2$,
or to disentangle the singlet,
nonsinglet and gluon distribution, or to determine the parity
violating structure
function $F_3$ (which only contributes to neutrino DIS). 
Knowledge of the next-to-leading order corrections to eq.~(\ref{scv})
allows then to fix the scale at which $\alpha_s$ is evaluated.

The current\cite{pdb} global average value of $\alpha_s$ from
unpolarized 
scaling violations is $\alpha_s(M_z)=0.117\pm0.002({\rm exp.})
\pm0.004({\rm th.})$. 
This average includes results from
global analyses of
electron\cite{vm}, muon\cite{nmcas} and neutrino\cite{ccfr} DIS
data, i.e. obtained by introducing a parameterization of the
singlet,
nonsinglet and gluon at a reference scale,
evolving it up to the values of $Q^2$ 
at which data are available, and inverting the Mellin transform in
order to compare the structure function $f(x,Q^2)$ to the data.
In all analyses included in the above average,
computations (and specifically the solution of
the evolution equations) 
are performed to next-to-leading order (NLO).   

The current value is to be contrasted with the
earlier\cite{pdbold}  value $\alpha_s(M_z)=0.112\pm0.002({\rm exp.})
\pm0.004({\rm th.})$. The central value has gone up because of a
re-evaluation of the CCFR result from neutrino scattering, which
now gives\cite{ccfr} $\alpha_s(M_z)=0.119\pm0.002({\rm exp.})
\pm0.004({\rm th.})$ (the previous central value was\cite{ccfrold}
$\alpha_s(M_z)=0.111\pm0.002({\rm exp.})
\pm0.003({\rm th.})$). The variation is mostly due to a
re-calibration of the energy of the neutrino beam along with smaller
improvements and corrections. Because of the smallness of its
error, the CCFR determination now dominates the value of $\alpha_s$
from scaling violations, and we will thus discuss it in some detail.

The peculiarity of the CCFR determination
of $\alpha_s$ which justifies its small statistical error is the
fact that the use of a neutrino beam allows a simultaneous
determination of the structure functions $F_2$ and $F_3$, and thus a
reliable separation of the singlet and nonsinglet components.
The range of $Q^2$ covered by the data is reasonably high,
$5\le Q^2\le 125$~GeV$^2$, and an estimate of the error associated 
to power corrections is included.   

A shortcoming of the analysis, however,
is the relatively crude form of the
parton parameterization adopted: no attempt is made to fit the
detailed
$x$ shape of the structure function; this may lead to
an underestimate of the systematic error.\cite{abfr} 
A perhaps more
interesting potential source of trouble is related to
higher order corrections, whose impact was not studied  
(the corresponding error 
is assumed to be the same as determined
in the unrelated analysis of ref.\cite{bcdms}).

In fact, the NNLO anomalous dimensions have been recently determined,
at least for even integer values of $2\le N\le 10$.\cite{nnlo}
Since the large $N$ (see Sect.~3) and small $N$ (see Sect.~4) behavior
of the anomalous dimensions is also known, it is possible  
to reconstruct
approximately the full anomalous dimension by interpolation.
A NNLO
analysis can then be attempted.
The CCFR  $F_3$ 
(i.e. pure nonsinglet) data have been recently
reanalyzed in this way.~\cite{kata,katanew} The inclusion of NNLO terms
results in a
decrease of the $\chi^2$ (by 
about 5 units with about 10 d.o.f.), and in
a lower  central value of $\alpha_s$ by $\delta \alpha_s(M_z)=
0.004$. Since 
this is equal to the total theoretical error quoted in
ref.\cite{ccfr}, it suggests that this error 
may actually
be underestimated; a careful inclusion of  other sources of
theoretical uncertainty 
(specifically related to heavy quark thresholds) suggests that
a more realistic estimate  might be
$\Delta[{\rm th., NLO}] \alpha_s(M_z)=0.006$, which at NNLO is 
reduced to
$\Delta[{\rm th., NNLO}] \alpha_s(M_z)=0.003$ (these numbers apply to
the pure nonsinglet analysis of ref.\cite{kata}).  
An alternative NNLO determination from ``world'' data~\cite{ynd} 
finds 
$\alpha_s(M_z)=0.1163\pm0.0016({\rm exp.})
\pm0.0016({\rm th.})$. Here,  the smaller statistical error is 
due to using
a wider dataset, while the smaller theory error comes from a 
less conservative treatment of
theoretical errors: a less general parameterization of higher twists is
used, and higher orders are estimated by different linearization of
the Altarelli-Parisi solution rather than by scale variation.

Another important potential source of error is related to
the inclusion of contributions to the evolution equations beyond
logarithmic accuracy, 
i.e. corrections to eq.~(\ref{scv}) suppressed by
powers of $Q^2$ (higher twist corrections). 
We will discuss this issue in detail in Sect.~3.
In ref.\cite{ccfr} an
estimate of higher twist corrections is included on the basis of a renormalon
model (see Sect.~3.1), and assigned a 100\% error, which gives
an uncertainty $\Delta({\rm HT, NLO}) \alpha_s(M_z)=0.001$. A (much)
more conservative estimate is obtained by parameterizing
the higher twist
corrections by a function $h(x)\over Q^2$ which is left 
as an extra free parameter for each $x$ data bin
in the fit. In such case the theory error
is enormously increased to\cite{kata}
$\Delta({\rm th.+HT, NLO}) \alpha_s(M_z)=0.010$, which can only be
reduced by performing a  NNLO analysis, in which case one gets
$\Delta({\rm th.+HT, NNLO}) \alpha_s(M_z)=0.004$. This error estimate
is perhaps a exceedingly conservative, since a very large number
of extra free parameters is being introduced (16 extra
parameters). This inevitably inflates the uncertainty unless the
statistics is very high; notice that this procedure was adopted in the
`classic' BCDMS  determination~\cite{vm} of $\alpha_s$.
However, there clearly  is a   correlation 
between higher twists, higher
perturbative orders and value of $\alpha_s$ (we shall come back to
this in Sect.~3.2), which  suggests
that higher corrections are not negligible to present accuracy.

The correlation between higher twist and higher orders raises an
interesting problem:\cite{ale} so far, 
different sources of uncertainty have always been assumed to
be uncorrelated, and added in quadrature. This assumption however
cannot be completely correct, either experimentally or
theoretically. 
On the experimental side, many sources of  systematics
are either common to all data in a given experiment or highly
correlated point by point, and simply determining the error on each
data point by adding statistical and systematic errors cannot be
right. The neglect of correlations not only leads to an incorrect
estimate of the total $\chi^2$ (which indeed is often found to be
unrealistically mall, i.e.  significantly
smaller than one per degree of freedom), but more seriously it could
bias the best-fit value of $\alpha_s$. 

Indeed, a recent reanalysis\cite{ale} of the
BCDMS data shows that the central value
shifts from $\alpha_s(M_z)=0.113$\cite{vm} to $\alpha_s(M_z)=0.118$
if experimental correlations are included.
Moreover, on the theoretical side it's clear that not all the
uncertainties discussed above are uncorrelated: for instance (see also
Sect.~3.2) if higher twist and higher order contributions are not
entirely independent, the associate uncertainties should not be simply
added in quadrature. Rather, if the higher twists are treated as free
parameters they should simply re-fitted as the scale is
varied. Proceeding in this way the result improves rapidly with the increase
in statistics. For instance, the large error found above remains essentially
unchanged  if one restricts the analysis to the CCFR $F_3$ data
alone\cite{katanew},  but it
decreases to
$\Delta({\rm th.+HT, NLO}) \alpha_s(M_z)=0.004$ when the
CCFR $F_2$ data are also included\cite{alkat}. Likewise, the error on the
BCDMS/SLAC NLO extraction decreases from\cite{vm}
$\Delta({\rm total}) \alpha_s(M_z)=0.003$ to\cite{ale}
$\Delta({\rm total}) \alpha_s(M_z)=0.0017$.

In conclusion, the value of $\alpha_s$ from unpolarized
scaling violations can be extracted from current data with excellent
statistical accuracy: the dominant source of error is theoretical,
related  to higher order corrections. Progress is now related to 
a satisfactory treatment of theoretical errors and their
correlations, and reliable NNLO calculations.

A  determination of $\alpha_s$ from scaling violations
of the {\it polarized} structure function $g_1(x,Q^2)$
has also been presented recently~\cite{abfr}. The result is
$\alpha_s(M_z)=0.120\epm{0.004}{0.005}({\rm exp.})
\epm{0.009}{0.006}({\rm th.})$. The statistical accuracy of this
result (which includes a detailed study of the error due to the
way parton distributions are parametrized) is justified by the
recent remarkable improvement in statistical accuracy and kinematic
coverage of the polarized data.\cite{polrev} Indeed, the error is
already dominated by the theoretical one: this is however due to the
fact that the available data are taken at relatively low $Q^2$
(the most accurate data have an average $Q^2$ of a few GeV$^2$), so
that higher order and higher twist corrections are very large. 
A competitive determination of $\alpha_s$ would require the
availability of polarized data at higher energy (such as could be
acquired at a polarized HERA\cite{polhera}).

A byproduct of these determinations of $\alpha_s$ is the determination
of the polarized and unpolarized gluon distribution, which is made
possible by the availability of structure function
data at several scales $Q^2$: 
according to eq.~(\ref{scv}),
knowledge of the nonsinglet and singlet 
structure function at two scales at least determines simultaneously
$\alpha_s$ and the gluon. Note that the increase in uncertainty 
on $\alpha_s$ due to the need to determine the gluon simultaneously
is by construction included in the statistical error, and thus
appears to be
subdominant. Scaling violations provide the most precise way to
determine the unpolarized gluon for $x\lsim 0.3$ (where it is known to
about 10\% accuracy at scales of a few GeV),\cite{gerr} and the only
available handle on the polarized gluon (which is only known for
values of $N\sim 1$ and then with an uncertainty greater than 
50\%)\cite{abfr,bfr}

\subsection{Sum rules}

Rather than using the full set of structure function data in the
$(x,Q^2)$ plane to infer the value of the strong coupling (and parton
distributions) from a global fit to the scaling violations of all
Mellin moments, it may be advantageous to concentrate on one 
single moment,
specifically the first moment of $F_3$ and $g_1$. The reason is that
these first moments are proportional to the nucleon expectation value
of a conserved operator, the vector and axial currents, respectively,
with a coefficient determined by the operator-product expansion.

The fact that the operator is conserved means that its
expectation
value is scale-independent (the anomalous dimension vanishes to all
orders), and equal to a conserved quantum number: the baryon number in
the case of $F_3$, and the axial charge $g_A$ (measured in
$\beta$-decay) in the case of $g_1$. The relation between the first
moments of $F_3$ and  $g_1$ and the respective charges are known as
Gross--Llewellyn-Smith and Bjorken sum rules, respectively.
The Wilson coefficients that relate the charges to
the first moments of the structure functions have been
determined up to N$^3$LO (i.e. up to $O(\alpha_s^3)$).\cite{nnnlo}
It follows that 
knowledge of the first moment of the structure
function at two or more scales allows one to check the normalization
of the sum rule, which is a parameter-free prediction of QCD, and its
$Q^2$ dependence, which depends on the value of
$\alpha_s$.

 Current tests of Bjorken the sum rule~\cite{abfr,elkar}  are in
excellent agreement with the QCD prediction. It is however more
interesting to assume the correctness of QCD and thus of the sum
rules, and use them to extract $\alpha_s$: in principle, a single
measurement of the moment at one scale provides then $\alpha_s$
 at that scale to 
N$^3$LO accuracy. It is therefore possible to keep under control and
greatly reduce the error related to higher order corrections, which is
dominant in the determinations discussed in Sect.~3.1.

This favorable state of affairs is spoiled by the difficulty of
attaining a reliable determination of 
the first moment of the structure function,
i.e. its integral over the full $x$ range. Since
the invariant energy $W^2$ diverges as $x\to0$, it is
obviously impossible to measure
down to $x=0$. However, it has been only recently realized
that the convergence of the integral at small $x$ is rather slow
in the presently accessible $x$ region. This slow convergence
is due to the fact that
both polarized and unpolarized nonsinglet quark distributions
(and thus, correspondingly, $g_1$ and $F_3$) appear to grow as
$1\over \sqrt{x}$ at small $x$ (in agreement with the unpolarized
Regge theory expectation, but in disagreement with the polarized
prediction of a power falloff of $g_1$ at small $x$).
 
\begin{figure}
\centering\mbox{
    \epsfig{file=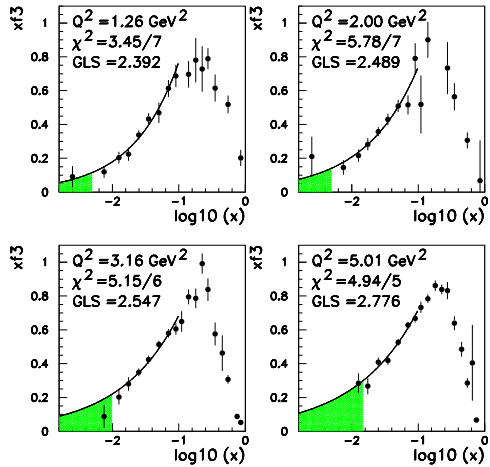,width=7cm}}
 \caption[]{The structure function $F_3(x,Q^2)$; GLS denotes the 
area under the
curve. The shaded area is the contribution to GLS from the small $x$
    extrapolation (from Ref.\cite{ccfrgls}).}
\end{figure}
The importance of the small $x$ extrapolation is highlighted in
Fig.~2: it is clear that the choice of a specific
functional form for
the small $x$ extrapolation greatly constrains the result for the
first moment, and can thus lead to an underestimate of the
uncertainty due to the small $x$ extrapolation.
The associated uncertainty has been studied in the polarized case in
Ref.\cite{abfr} by explicitly varying the functional form of the
extrapolation. As a result, the value of $\alpha_s$ extracted from the
Bjorken sum rule acquires a very large uncertainty $\Delta
\alpha(M_z)=\epm{0.010}{0.024}$: the value includes statistical and
extrapolation errors, but is dominated by the latter. 
This large error overwhelms any advantage which may be obtained from
the accurate perturbative knowledge of the Wilson coefficient, 
and makes the
extraction of  $\alpha_s$ from it phenomenologically not viable with
present data.
In the
unpolarized case a much more optimistic estimate ($\Delta
\alpha(M_z)=\epm{0.006}{0.007}$) 
was obtained\cite{ccfrgls} by assuming a power-like behavior and
varying the exponent of the power. In the unpolarized case, however,
the issue is largely academic, since the need to use neutrino data
to determine $F_3$ introduce several other sources of large
systematic error, specifically related to the normalization of the
cross section and the estimate of the charm contributions.

In conclusion, the extraction of $\alpha_s$ from DIS sum rules is
theoretically very clean and could be a nice laboratory to study
the convergence of the perturbative expansion of the leading twist
Wilson coefficients and its relation to higher twist
terms.\cite{elkar} It is however not viable from a
phenomenological point of view at present, mostly
due to the insufficient
kinematic coverage of the data which does not allow a reliable
determination of the first moment of the structure functions.

\subsection{Scaling violations at small $x$}

Perhaps the most surprising experimental
discovery at HERA is that the very strong scaling violations
predicted by
perturbative QCD at small $x$ are actually seen in the data,
rather than being
cut off by some nonperturbative mechanism: in the
HERA kinematic range $F_2(x,Q^2)$ rises at small $x$ in accordance
with the QCD prediction,\cite{dgptwz} 
rather than displaying the set-in of a
nonperturbatively generated Regge-like behavior, which many expected
and pre-HERA data seemed to suggest.\cite{oldnmc}  
It is natural to think that if scaling
violations are strong, then  it must be possible to determine
$\alpha_s$ accurately. 

In fact, this turns out to be case also because
of more subtle theoretical reasons. Indeed, the small $x$ rise of
structure functions in perturbation theory is due to the fact that
the gluon anomalous dimension has a pole in the $N$ plane at $N=1$.
Solving the evolution equations it is easy to see that
this pole translates into a rise of the gluon
distribution and thus the structure function\cite{dgptwz}
$F_2\sim\alpha_s xg(x,Q^2)\sim\alpha_s(Q^2)
\exp[2 \gamma \sqrt{\xi\zeta}]$ (with $\xi=\ln{1\over x}$, $\zeta=
 \ln{1\over \alpha_s(Q^2)}$ 
and $\gamma={12\beta_0}$), stronger than any power
of $\ln {1\over x}$ but weaker than any power of $x$
(poles at smaller values
of $\Re N$ give contributions suppressed by powers of $x$ at small
$x$).

Expanding the
anomalous dimensions about the pole leads to a systematic expansion
of the small $x$ structure function, where the $k$-th subleading 
order is
suppressed by  $\exp[\zeta/\xi]^{k/2}$.\cite{das} The relevant point here is that truncating this
expansion leads to an evolution equation (which is the small $x$
approximation to the Altarelli-Parisi equation) which is local
in $x$ and $Q^2$; in particular, at leading order the gluon evolves
according to a simple wave equation with a negative mass,
while the (singlet) quark
$q(x,Q^2)$ is
determined by the gluon:
\bea
{\partial^2 \left[x g(x,Q^2)\right]
\over \partial \xi\partial\zeta}&=&\gamma^2
x g(x,Q^2)\\
\label{weq}
{\partial \left[x q(x,Q^2)\right]
\over \partial \zeta}&=& {n_f\over 9} \gamma^2
x g(x,Q^2).\label{qeq}
\eea

\begin{figure}
\centering\mbox{
    \epsfig{file=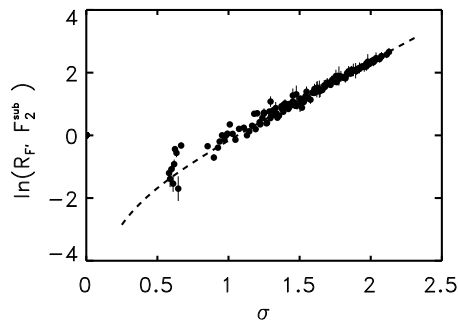,width=7cm}}
 \caption[]{Double scaling plot of the HERA data published up to
    1996. The dashed line is the universal (parameter-free)
    prediction, computed with $\alpha_s(M_z)=0.114$. (From Ref.\cite{lech}).}
\end{figure}
It follows that
the leading asymptotic behavior of the structure function
at small $x$ is a rise in the product of the two
variables $\xi$ and $\zeta$. The scale of the rise is set by
$\alpha_s(Q^2)$ (through the definition of $\zeta$). 
The structure function depends only on the combination
$\sigma\equiv\sqrt{\xi\zeta}$, and the asymptotic dependence on
$\sigma$ 
is then universal (double asymptotic scaling\cite{das}).
It is important to notice that the asymptotic slope of this rise is a
universal prediction: since asymptotically the
running of $\alpha_s$ is driven by the leading order $\beta$ function
and independent of the value of $\alpha_s$ (i.e. of $\Lambda$), the
asymptotic slope of $F_2$ is a parameter-free prediction of QCD, and
its experimental determination provides a basic test of
the theory~\cite{test}. In fact, this is a more fundamental test of the QCD
dynamics  than those provided
by the sum rules discussed in Sect.~2.2: the value of the slope of the small
$x$ rise is fixed entirely by the Casimir operator of the gauge
group and the number of quark flavors, whereas the values of the CCFR and
Bjorken sum rule are determined, respectively, by the quark baryon number 
and by the nucleon $\beta$-decay constant.

However, for the sake of phenomenology it is again more
convenient to assume the correctness of QCD evolution, and thus of this
asymptotic prediction. In the region where the asymptotic prediction
holds, but two-loop running effects are still relevant, it is then possible
to extract $\alpha_s$ from the observed scaling violations. 
Such a  determination appears to be particularly favorable, since
these scaling violations are strong, and
independent of the form of the parton 
distributions: in particular, one does not need an independent
determination of the gluon distribution since the quark and gluon
distribution are asymptotically proportional. 
There is however a major caveat, namely, whether 
one can trust NLO QCD evolution in the first place.
This is not obvious, essentially because higher order
corrections to the Altarelli-Parisi anomalous dimensions are expected
to induce a yet stronger rise: the related problems will be discussed
in Sect.~4. Also, one may think that eventually the
perturbatively generated rise should be cut off by a nonperturbative
mechanism. 

A simple way of making sure that NLO evolution applies is
to test for double asymptotic scaling
(after dividing out universal
subasymptotic corrections\cite{zako,lech}), to which it reduces at
small $x$. Such a comparison is
shown in Figure 3. Notice that the asymptotic prediction 
only depends on the value of $\alpha_s$: 
a description based on  NLO evolution thus appears adequate. 

Of course
an accurate determination requires comparison to the full NLO
prediction (rather than just the asymptotic behavior). Such 
a determination was done in Ref.\cite{alphaold,alphanew}, on the
basis of the data then available, with the result
$\alpha_s(M_z)=0.120\pm{0.005}({\rm exp.})
\pm{0.009}({\rm th.})$. On top of the advantages already discussed
(strong scaling violations, little dependence on 
the input parton distributions) it also 
 turns out~\cite{alphanew} that higher twist
corrections are negligibly small in the HERA kinematic region (a
result which recent data have confirmed\cite{mrstht}). 
The relatively
large experimental
error is due to the poor quality of the data then available,
which also justified 
a crude treatment of the
experimental systematics. The large theoretical error is mostly due to
uncertainty related to the possibility of deviation from NLO evolution
at small $x$, and to factorization and renormalization scale
uncertainties: these could be both reduced if
$\alpha_s$ were extracted from the current data 
(Fig.~1) which
are much more precise and have a wider kinematic coverage.

\begin{figure}
\centering\mbox{
    \epsfig{file=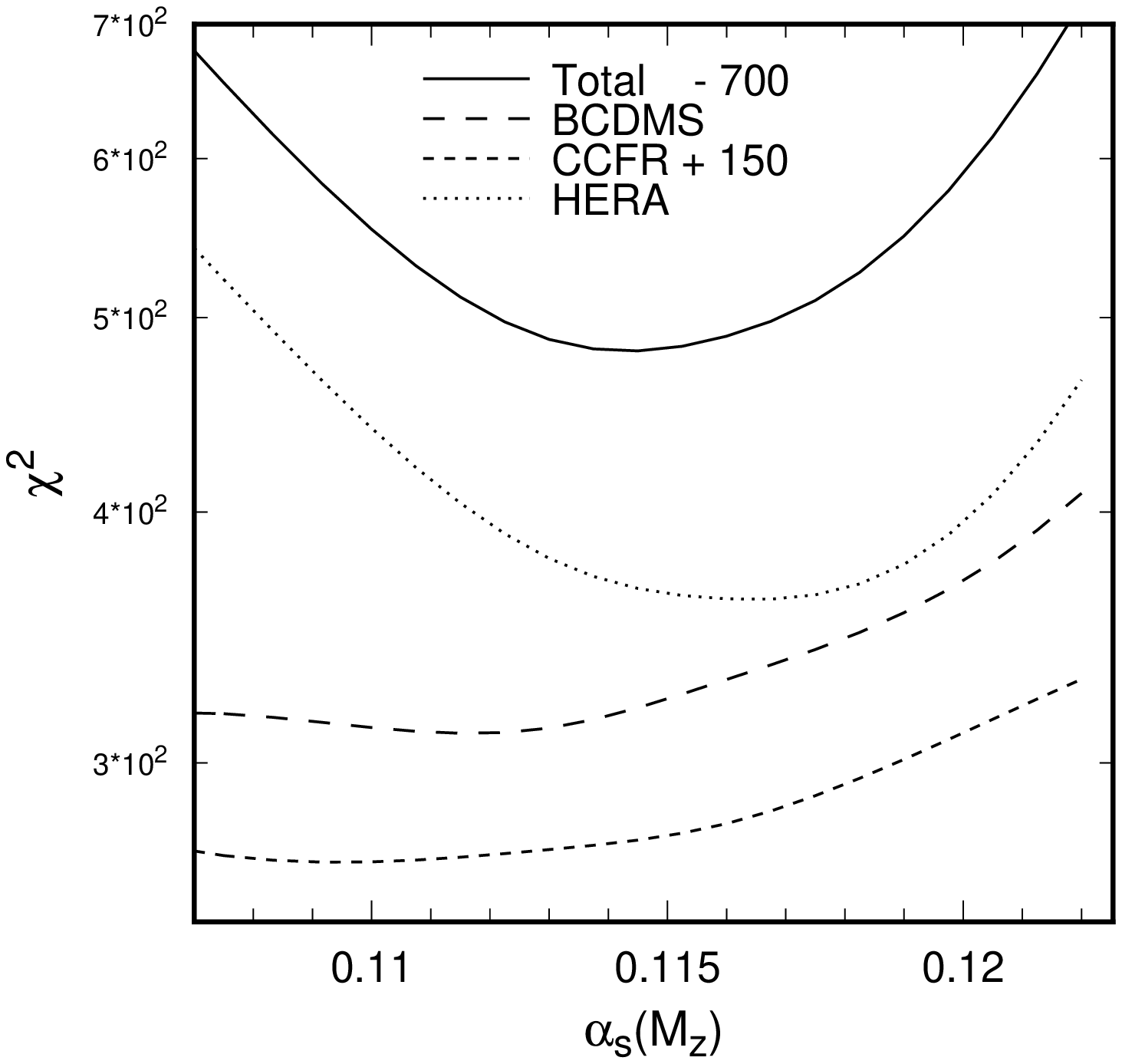,width=4cm}
\epsfig{file=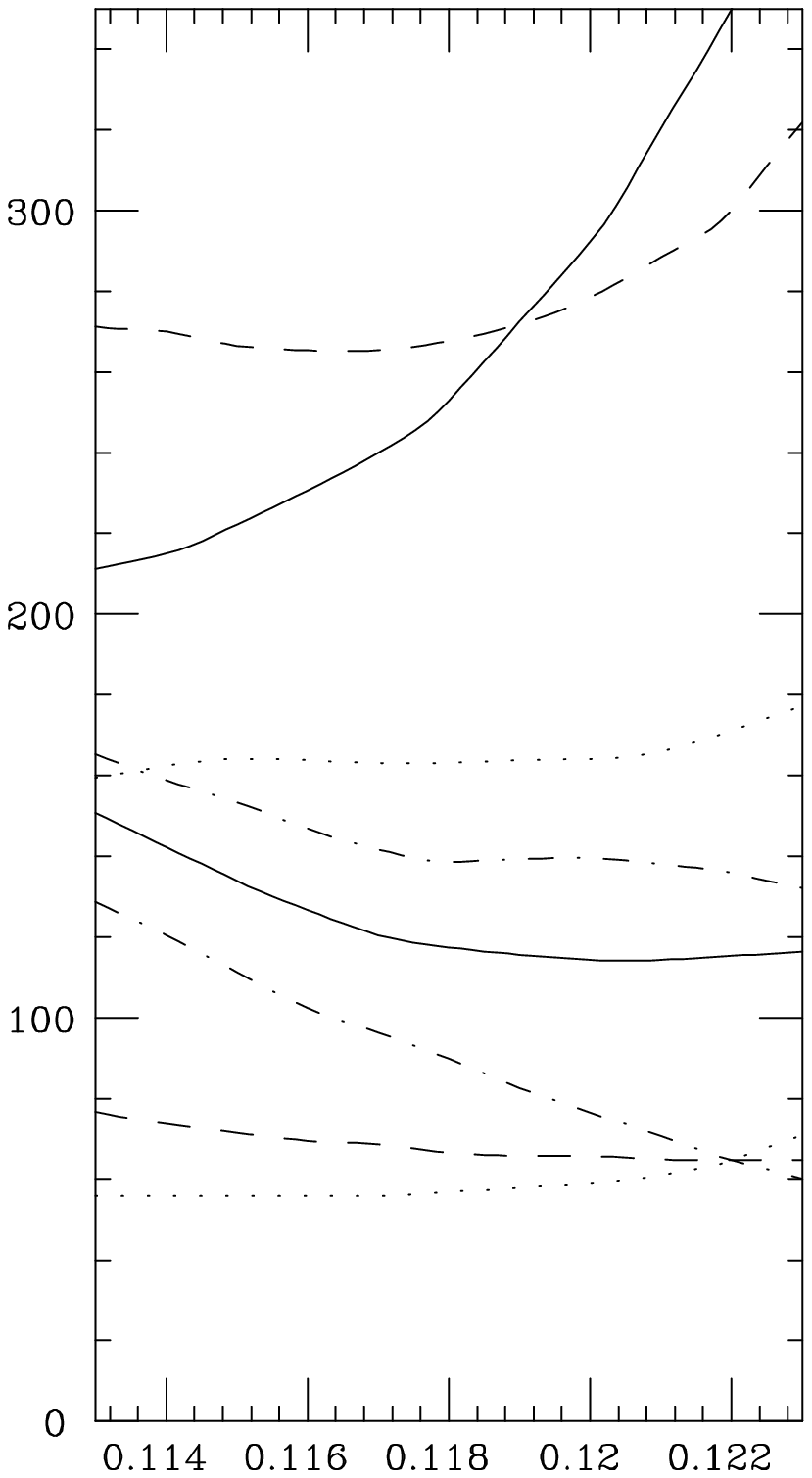,width=6.5cm}}
 \caption[]{Contribution to the $\chi^2$ of the fit
from various datasets for the global fits of Ref.\cite{cteq} (left)
and Ref.\cite{mrst} (right, adapted from the original ref.). 
In the latter case the curves,
from top to bottom on the left denote the contributions
from ZEUS, BCDMS, NMC, H1, SLAC, CCFR $F_2$, CCFR $F_3$ and E665.}
\end{figure}
The advantage of an extraction of $\alpha_s$ at small $x$
can be seen by comparing the contributions to the $\chi^2$ of a
global  fit of parton distributions to DIS data from various datasets
as a function of $\alpha_s$ (Figure 4).\footnote{The 
CTEQ fit\cite{cteq} uses the 
HERA (i.e. H1+ZEUS) collected up to 1994 and the old\cite{ccfrold}
CCFR data, while
the MRST fit\cite{mrst} includes HERA data
up to 1995 and the new CCFR data\cite{ccfr};
the BCDMS\cite{bcdms}
data are the same in the two fits.}
Of course, a global fit doesn't necessarily provide a
good determination of $\alpha_s$ for each of the separate datasets
because of the difficulty in combining the data from different
experiments and in different kinematic region (and in particular 
their
uncertainties). So,
for instance, the BCDMS data, which give an excellent\cite{vm}
determination of $\alpha_s$, would appear to have no minimum in the
fit of Ref.\cite{mrst}. Yet it is clear that the minimum in the 
small--$x$
data always
is more stable, deeper, and self-consistent than any of the other
datasets. A determination of $\alpha_s$ from the data shown in Fig.~1
is highly desirable and could be one of the most precise on the market.

\section{Large $x$ and power corrections}
Large values of $x$ correspond to the kinematical region where the
invariant mass of the final state $W^2$ eq.~(\ref{kin})
is small. In this region power corrections
proportional to either $\Lambda^2\over W^2$ or $M^2\over W^2$
(where $M$ is the hadron target mass) are potentially large. 
Target mass corrections are of purely kinematical origin, and are
thus completely understood from a theoretical
viewpoint, and routinely included in current unpolarized
data analyses. Note, however, that in the polarized case 
the computations necessary in order to implement
the relevant phenomenology have been performed only
recently.\cite{ucc} Power corrections in $\Lambda^2\over W^2$
can be 
generated by the resummation of
logs of $1-x$ which are present at all orders in perturbation
theory. 

\subsection{Leading $\ln(1-x)$ and power corrections}

Logs of $1-x$ are generated  in the perturbative computation
of the DIS cross section 
for kinematic reasons: as
$x\to1$ the phase space for the emission of an extra parton  is
suppressed, so that each extra emissions carries a factor of
$\alpha_s(Q^2) \ln(1-x)$. At the leading logarithmic level, these
contributions are resummed\cite{llx} by replacing
\be
\as(Q^2) \to\as((1-x)Q^2)={\alpha_s(Q^2)\over
1+\beta_0\alpha_s(Q^2)\ln(1-x)}  
\label{rep}
\ee
 and bringing the coupling
inside the $x$ integration when evaluating the Mellin 
moments of the splitting function on the r.h.s of
the evolution equations eq.~(\ref{scv}). This means that the coupling
is effectively evaluated at the scale $W^2$ eq.~(\ref{kin}); 
as $x\to1$, $W^2\to1$ 
and 
the growth of the coupling in the infrared generates a factorial
divergence in the perturbative expansion, which is in turn related to
a powerlike 
correction to the structure function. 

This can be understood by
considering the, say, first Mellin
moment of $\alpha_s$ after the substitution eq.~(\ref{rep}):
\bea
I[\alpha_s]\equiv\int_0^1\! dx \alpha_s((1-x) Q^2) &=&
\alpha_s(Q^2) \sum_{k=0}^\infty
(-1)^k [\beta_0\alpha_s(Q^2)]^k  \int_0^1 \! dx\, \ln(1-x)^k 
\nonumber\\
&=& \alpha_s(Q^2) \sum_{k=0}^\infty k! (\alpha_s(Q^2)\beta_0)^k .
\label{aser}
\eea
The factorially divergent series can be summed by Borel resummation,
i.e. by defining 
\be
I[\alpha_s]\equiv\int_0^\infty\!ds\,e^{-s/\alpha_s(Q^2)}
B[s].
\label{borel}
\ee
It is easy to see that $B[s]={1\over 1-s\beta_0}$.
Thus, if we want to determine $I[\alpha_s]$ from eq.~({\ref{borel}) we
must integrate over the pole at $s=s_0\equiv{1\over\beta_0}$. This generates a
contribution to  $I[\alpha_s]$ proportional to
$e^{-s_0/\alpha_s(Q^2)}={\Lambda^2\over Q^2}$. Reconstructing the full
$x$ dependence from the computation of all Mellin moments shows that
the power suppressed contribution is
actually proportional to $\Lambda^2\over W^2$,
i.e. rapidly growing at small $x$. 

The strength of this $\Lambda^2\over Q^2$
contribution is ambiguous however: for instance,
its sign depends on
whether the path of integration along $s$ is taken above
or below the pole.
The ambiguity
is to be expected  since we have obtained the result from a
leading twist computation, i.e. not including terms suppressed by
powers of $Q^2$,
and can only be fixed by performing the computation to power
accuracy. Of course, as long as we are interested in the leading
twist, leading $\ln (1-x)$ result, we may simply refrain from
expanding out 
$\alpha_s((1-x)Q^2)$. 

This suggests two directions of progress in the
description of the large $x$ region. On the one hand, we may push the
resummation of $\ln(1-x)$ contributions beyond the leading log
level. Indeed, the next-to-leading resummation is also
known;\cite{nllx} and a systematic method to resum subleading
contributions to all orders has been suggested 
recently.\cite{lxope} This determines
the  large
$N$ behavior of the anomalous dimensions and coefficient functions.
On the other hand, we may use the fact that
the all-order summation of leading twist contributions generates
power-suppressed terms as a means to acquire information on the higher
twist terms in the operator-product expansion. Considerable progress
has been made recently in the latter direction by means of the
so--called renormalon method.\footnote{It is impossible to cover here
the very rich
literature on this subject. The
reader is referred to the excellent recent review Ref.\cite{benrev}.
}  

\subsection{Renormalons and phenomenology}
Because the structure function is a physical observable,
the ambiguity proportional to $\Lambda^2\over Q^2$ generated by
the Borel resummation of classes of leading twist
contributions must be cancelled by an equal and opposite 
ambiguity in the next-to-leading twist
contribution. Indeed, such an ambiguity
arises when regulating the power ultraviolet divergence of the
next-to-leading twist, i.e., the pole in the Borel transform must
cancel between leading and next-to-leading twist. We can then take the
coefficient of the pole in the leading twist as an estimate
of the full next-to-leading twist computation. The rationale for this
is the same as when estimating a NNLO corrections by
varying the factorization scale in a NLO computation. Of course,
 we would
expect this estimate  to be only qualitatively correct.
 
A specific class of factorially divergent contribution, related to the
small momentum region of integration on loop momenta (infrared
renormalons) can be isolated and computed for a variety of
observables, and may be used to this purpose.\cite{benrev}
In the case of DIS, the renormalon contributions
to the coefficient functions which relate structure functions to 
parton distributions have been computed both in the 
nonsinglet\cite{sfren} 
and singlet\cite{rensing} sector (the contributions to the nonsinglet
splitting functions are also known\cite{ugo}, but have not been
used for phenomenology). One can then convolute
this $\Lambda^2\over Q^2$ contribution to 
the coefficient function
with parton distributions taken from a NLO, leading twist fit of
structure functions data, and view the result
as an estimate of the $O\left({\Lambda^2\over Q^2}\right)$ part of the
structure function itself, up to an unknown coefficient. The result
can be compared to available phenomenological fits,\cite{vm}
where  
the structure function
is parameterized as
\be
F_2(x,Q^2)=F_2^{\rm l.t}(x,Q^2) +{h(x)\over Q^2},
\label{htpar}
\ee
and the function $h(x)$  is treated as a
free
parameter for each $x$ data bin. 
 
After adjusting the
coefficient (to about  three times of what would be obtained 
by Borel transform integrating below the pole, as discussed in 
Sect.~3.1)
the agreement with the data appears to be excellent: in particular,
the shape (as a function of $x$)
of the
observed twist four (i.e. $O\left({1\over Q^2}\right)$) 
contribution to $F_2$
appears to agree very well with the renormalon estimate.
This poses a
problem, since quantitative agreement is found, whereas only a
qualitatively correct description was expected. The result is
especially surprising in view of the fact that higher twist
contributions correspond to operators which measure multiparton
correlations in the target and should thus be target-dependent,
whereas the  $O\left({1\over Q^2}\right)$
contribution computed from renormalons is
universal, since it comes from  a coefficient
function.

\begin{figure}
\centering\mbox{
    \epsfig{file=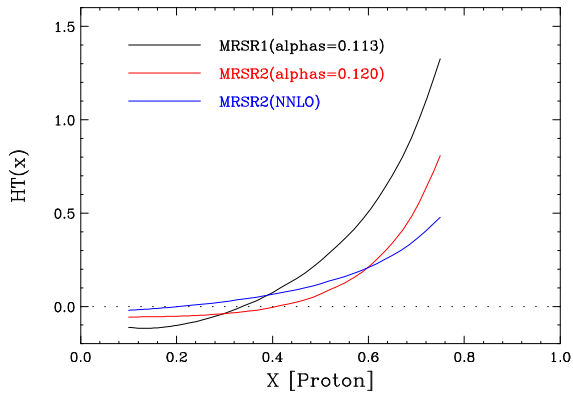,width=7.5cm}}
 \caption[]{Higher twist contribution eq.~(\ref{htpar})
to $F_2$ from the fit of
    ref.\cite{bodek}. The curves are labelled from top to bottom on
    the right.}
\end{figure}
A hint on the possible origin of this state of affairs comes from
a  recent phenomenological analysis\cite{bodek} of the
BCDMS\cite{bcdms}, SLAC\cite{slac} and CCFR\cite{ccfr} $F_2$ data
(see Fig.5). The fitted higher twist contribution agrees
well with previous fits,\cite{vm} even though here
a different\cite{mrs} set of
parton distributions and a wider dataset
are used. However, if the value of
$\alpha_s$ used in the computation of the leading twist part 
is changed from the low value used in Ref.\cite{vm} to a larger
value, a substantial reduction of the higher twist
contribution is observed. If $F_2$ is computed by
folding the NLO partons with the NNLO
coefficient function\cite{zij} the higher
twist is further reduced. A decrease of the fitted higher twist
contribution as
more perturbative orders are included has also been found
in
an analysis of the CCFR $F_3$ data;\cite{katanew} while the strong
anticorrelation between the value of $\alpha_s$ and the size of the
higher twist contributions has been observed in a recent reanalysis of the
BCDMS/SLAC data\cite{ale} and in a phenomenological extraction of
higher twists from ``world'' data.\cite{liuti}

This means that the fit doesn't really 
distinguish between log and power 
behavior:
 the higher twist correction 
extracted from the fits is in large part a parameterization
of missing leading twist contributions,
because of the truncation to NLO, or because the value of $\alpha_s$
is too small.
This explains naturally the success of the renormalon calculation,
which
is equal to the  resummation
of higher order contributions, and thus  provides a good
approximation to such missing
higher order terms. Notice that the leading
$\ln(1-x)$ terms discussed in Sect.~3.1 
are included in the renormalon estimate.
But this also means that the ``dynamical''
(i.e. target-dependent) part of the higher twist, which is unrelated
to higher perturbative orders and would spoil this agreement,
must be subdominant
(i.e. the higher twist is ``ultraviolet dominated''\cite{uvdom}). 
 In principle, it is alternatively
possible  
that dynamical higher twists have a similar 
functional form as the renormalon higher twist, and are thus not
easily disentangled from them.
This is  presumably true in the $x\to 1$ limit, 
in that a $1\over 1-x$ behavior is expected to be
generic on kinematic
grounds, and it indeed
happens  if dynamical higher twists
are e.g. estimated in a simple model based on free-field
theory.\cite{htmod}
It seems rather unlikely in general though, and it appears more
natural to simply conclude that dynamical higher twists are smaller,
and hidden by the renormalon contribution.

All this  has also interesting implications for the
leading twist phenomenology: it implies that the value of
$\alpha_s $ extracted from log scaling violations
can be underestimated  if  
higher twists are large in the fit  and, more importantly, 
the error on
$\alpha_s$ can be underestimated if higher twists are not
allowed to vary enough. 
These problems should be mitigated in a determination of $\alpha_s$
at small $x$,
where $W^2$ is large and indeed higher twists appear 
to be  small.\cite{alphaold,mrstht}

\begin{figure}
\centering\mbox{
    \epsfig{file=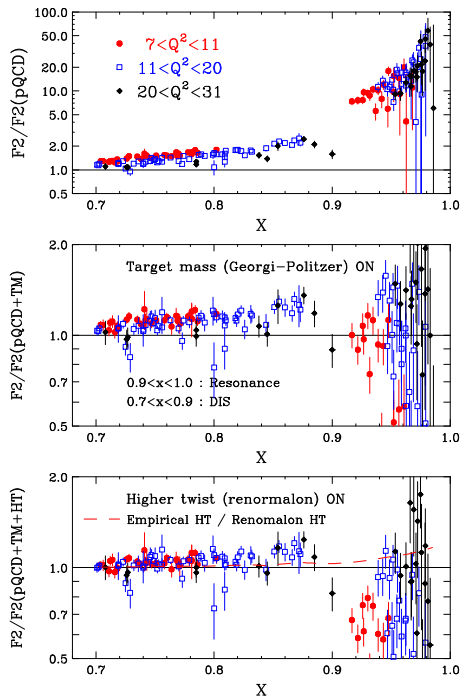,width=6cm}}
 \caption[]{Comparison of the $F_2$ data with the NLO calculation
    based on the partons of Ref.\cite{mrs}, without power corrections
    (upper), with target mass corrections (middle), with target mass
    and renormalon corrections (from 
    ref.\cite{bodek}). ``Empirical HT'' 
refers to a model functional form
    for $h(x)$ eq.~(\ref{htpar}).}
\end{figure}
One is thus
led to conclude that there is a hierarchy in power corrections
to structure functions at large $x$ (Fig.6): 
target mass corrections are
characterized by a relatively large scale of order of the target mass
$M\sim 1$~GeV;  higher twists due to the resummation of higher
orders in the \MS\ scheme have a characteristic scale of several
hundreds of MeV; and ``dynamical'' higher twist corrections appear
at a yet smaller scale.
 Even though there is no understanding of this subdominance, it
is not necessarily problematic: one would expect the natural scale of
dynamical higher twists to be of order $\Lambda$, and 
current data are compatible with this. 
If however the scale turned  be much smaller,
this would pose an
interesting dynamical problem: naive perturbation theory
would appear to work at low $Q^2$ better than it ought to. Be that as
it may, the phenomenological extraction\cite{simula} of the dynamical
higher twist corrections, which would give access to parton
correlations in the nucleon, appears to be a very difficult challenge
which will require new experimental information.
\section{Structure functions at small $x$}
As we discussed in Sect.~2.3, the rise of $F_2(x,Q^2)$
at small $x$ observed at HERA is in spectacular agreement with the
prediction obtained by approximating the NLO
anomalous dimensions with their highest
rightmost singularities  in
$N$-space (leading singularities, henceforth). 
This immediately raises a problem:   higher order (in $\alpha_s$)
contributions to thr anomalous dimensions contain higher order poles,
associated to 
higher order powers of $\ln {1\over x}$. One has thus  to
face the problem of summing these contributions. Because
at small $x$ according to eq.~(\ref{kin})
$W^2\approx {Q^2\over x}$, this is related to the determination of the
leading high energy behavior of parton-parton scattering cross
sections. 
\subsection{Altarelli-Parisi equations and summation of $\ln {1\over
x}$}
While solving the Altarelli-Parisi sums leading logs of $Q^2$,
the inclusion of  suitable higher order terms in the anomalous dimension
allows one to also sum other logs: for instance we discussed in
sect.~3.1 how to sum leading logs of $1-x$. The leading singularities
of the anomalous dimensions in the gluon sector have the generic
form\cite{lgad}
\bea
\gamma^{ij}=\sum_{k=0}^\infty \alpha_s^k \gamma^{(k)\,ij}
\label{lsexp}\\
\gamma^{(k)\,ij}=\sum_{n=1}^\infty a_n^{(k)\,ij} \left(C
{\alpha_s\over N-1}\right)^n\label{asexp}
\eea
where
$C={12\log2\over \pi} $ and  the coefficients satisfy
$\lim_{n\to\infty}
{a^{ij}_{n+1}\over a_n^{ij}}=1$. At leading order
in $\alpha_s$, the
only nonvanishing coefficients 
are in the gluon sector,
where they satisfy $a_n^{(0)\,gq}={4\over 9} a_n^{(0)\,gg}$
(with $n\ge1$), and can be
extracted\cite{lgad}  from
knowledge of the leading log high energy behavior of QCD perturbative
cross sections, which is in turn found by solution of the BFKL
equation.\cite{vddrev} If such contributions are included in the
anomalous dimensions, the solution to the
Altarelli--Parisi equation contains all leading logs of $1\over x$,
because a $k$-th order pole in $N$ space corresponds to a $\ln{1\over
x}^k$ contribution.

Naively, one may think that since the series
in eq.~(\ref{asexp}) has unit radius of convergence,
one must resum all orders in $\ln{1\over x}$ whenever the
expansion parameter
$\Xi(x,Q^2)\equiv
C\alpha(Q^2)\ln{1\over x}\sim 1$. This
is problematic because $C$ is very large so $\Xi(x,Q^2)
\gg1$
in the HERA region, and in fact $\Xi(x,Q^2)>1$ also in most of
the SLAC and all of the NMC region (see Fig.~7):  the small $x$ region
is outside the radius of
convergence of the series. This, in the pre-HERA 
age, led to the conclusion that the perturbative expansion
eq.~(\ref{asexp}) breaks down
at small $x$.\cite{levwr} 

This conclusion is however unwarranted,
because the quantity of direct physical interest is the
Altarelli-Parisi splitting function, related by inverse Mellin
transformation to the anomalous dimension
eq.~(\ref{lsexp}-\ref{asexp}). 
Now, the
inverse Mellin transform is essentially the same as the Borel transform
discussed in Sect.~3.1, so if the anomalous dimension 
as a series in $\alpha_s$ eq.~(\ref{asexp})
has a finite
radius of convergence,
the splitting function has an infinite radius of
convergence for all finite $x$. It follows that  
the leading $\ln {1\over x}$ summation converges for all
$x$, and can thus be used down to arbitrarily small $x$: 
in each
$(x,Q^2)$ region, it is sufficient to sum the first $k$ terms in
eq.~(\ref{asexp}),
where $[\Xi(x,Q^2)]^k/k!\sim1$, i.e. (from Stirling formula)
$k\approx 2.7 \Xi$ for $\Xi\gsim
1$. 
  Notice the difference in comparison to the large
$x$ case, discussed
in Sect.~3.1, where 
the leading large $N$ anomalous
dimension diverges factorially, so the
 leading $\ln(1-x)$ series has
a finite radius of
convergence (cfr. eq.~(\ref{rep})), one must resum an infinite
number of terms, and the resummation breaks down when $x\to1$.

In fact, it can be
shown\cite{afp} that the Altarelli-Parisi equation with
leading-singularity anomalous dimensions $\gamma^{(0),\,ij}$
eq.~(\ref{lsexp})
is completely equivalent, up to higher twists,
to the BFKL equation (in the inclusive case), 
which describes the leading log high energy asymptotics.  It follows that
one may proceed in two equivalent ways: either one determines the LO
coefficients $a^{(0)\,ij}$ eq.~(\ref{asexp}) from the BFKL kernel,
and then one inverse-Mellin transforms
the series eq.~({\ref{asexp}) term by term, thereby obtaining a
convergent series which may be truncated at finite order.\cite{summ}
Or, alternatively, one constructs ``nonperturbatively'',
i.e. numerically the leading-singularity anomalous dimension as a
function of $a/N$ directly from the BFKL kernel.~\cite{ehw}

Either way, one is lead to a
perturbative expansion of the
small-$x$ splitting function
\bea
P^{ij}(x,Q^2)=P^{(0)\,ij}+\alpha_sP^{(1)\,ij}+\dots\nonumber\\
P^{(p)\,ij}(x,Q^2)={1\over x}\sum_{n=0}^\infty a_{n+1}^{(p)\,ij} \left(C
{\alpha_s\ln{1\over x}\over n!}\right)^n.
\label{sfexp}
\eea
In the quark sector, the coefficients $a^{qj}_n$ 
start at
$O(\alpha_s)$ 
and are thus subleading
compared to the gluon ones; they are however also
known.\cite{nlqad} It is thus possible to determine the splitting
functions $P^{ij}$ to leading nontrivial order, add
them on top of usual NLO
splitting function (with suitable matching conditions),
and 
compare the result to the HERA data\cite{summ,ehw}. 

\begin{figure}
\centering\mbox{
    \epsfig{file=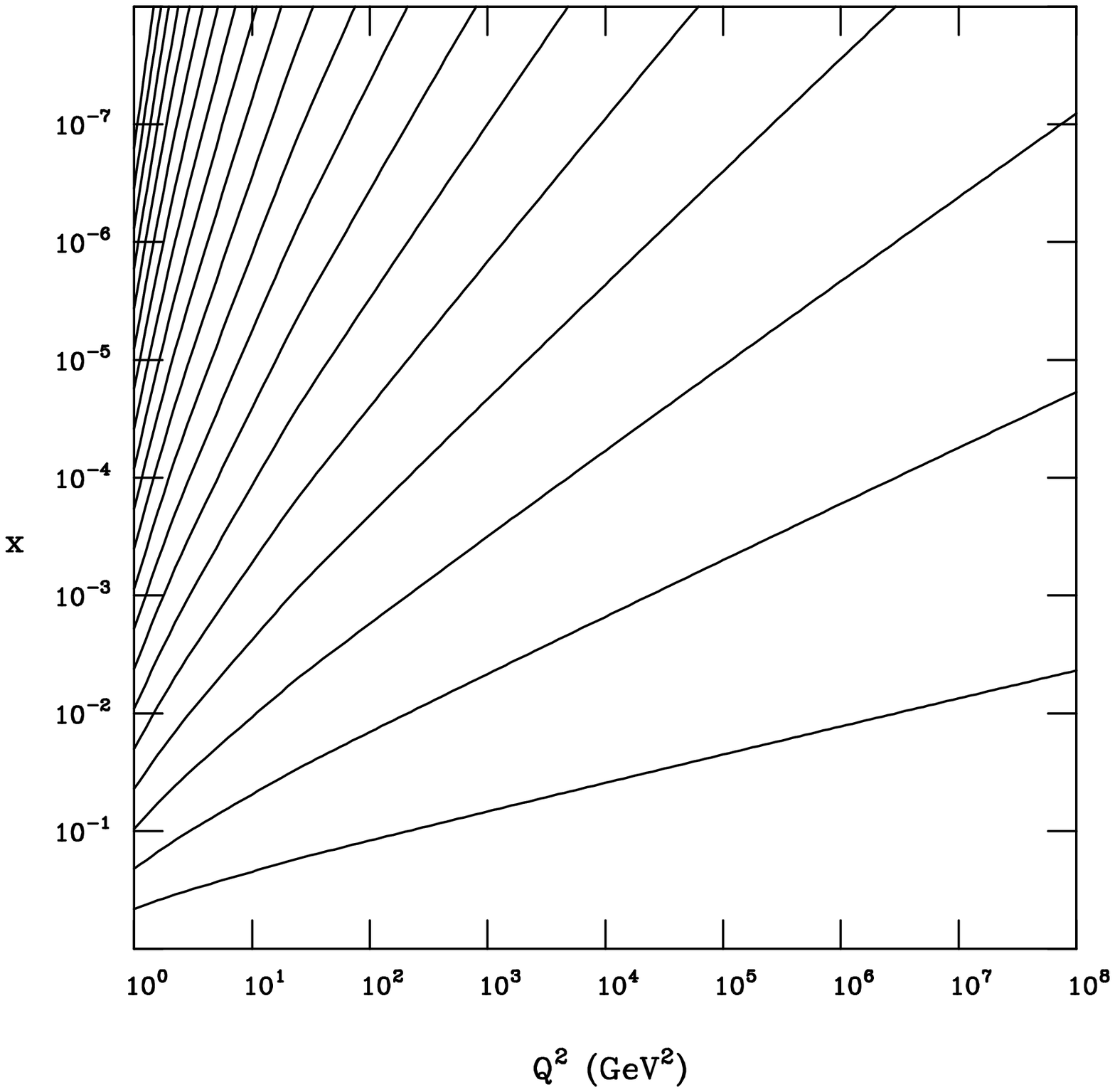,width=6.cm}\epsfig{file=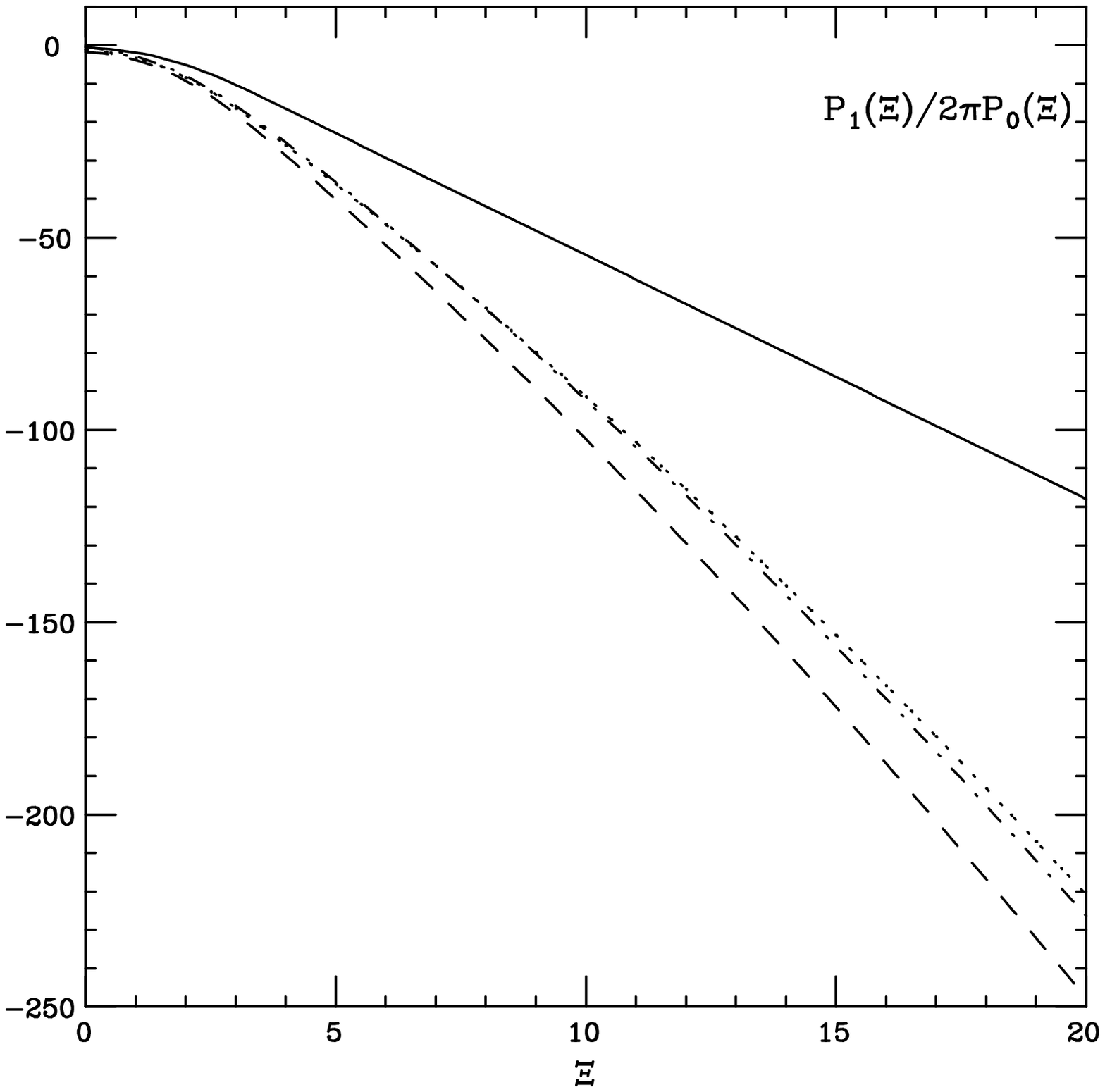,width=6.cm}}
 \caption[]{(Left) Contours of constant
    $\Xi\equiv C\alpha_s(Q^2)\ln({1\over x})
=1,2\dots 20$ (from bottom
    to top) in the $(x,Q^2)$ plane. 
as a function of $x$ and $Q^2$.
\quad (Right) 
The ratio of NLO to LO splitting functions eq.~(11) in various
    factorization schemes: \MS (solid), $Q_0$\cite{ciaqz} (dotted),
    ``physical''\cite{phys} (dashed and dot-dashed) (from Ref.\cite{sxap}).}
\end{figure}
Unfortunately, the results of
this comparison are discouraging: as long as the HERA data were still
relatively scarce, no evidence of the $\log{1\over x}$ summation could
be seen,\cite{summ} but as soon as they became accurate enough, the
presence of such effects could be excluded to very high
accuracy, essentially because they induce much stronger scaling
violations than observed in the data.\cite{roma}
In fact, the only way current data can be
reconciled with evolution equations 
that include a summation of small
$x$ effects is by artificially
fine--tuning the factorization scheme in such a way
as to remove these effects from the measured region.

A possible way out
of this {\em impasse} consists of exploiting the wider factorization
scheme freedom which is related to the summation of small-$x$
contributions: for instance, one can always reabsorb any undesirable 
scaling violations induced by the leading-order contributions
in eq.~(\ref{lsexp}), which only appear in the gluon sector, in a
redefinition of the gluon normalization, at the only cost of changing
the next-to-leading coefficients $a^{(1)\,gg}_n$. 
It  could then be that the effects which cause
disagreement with the data are a spurious consequence of the
choice of  factorization
scheme, and would disappear in a ``physical'' scheme where each
parton distribution is identified with a physical
observable.\cite{phys} 
In order to check
whether this is the case, however, the full set of NLO coefficients
$a_n^{ij}$ eq.~(\ref{asexp}) is needed: one may then
verify whether in this scheme the
undesirable scaling violations actually
disappear.

Thanks to the  recent calculation of the
next-to-leading corrections to the high--energy QCD 
asymptotics,\cite{fl} 
which allows a
determination of the next-to-leading coefficients
$a_n^{(1)\,gg}$,  it is now possible to resolve this issue.
The outcome of the calculation, however, brought a new surprise:
the NLO coefficients are negative and grow very large\cite{brus,blum} 
in comparison
to the leading order ones  
as the order $k$
in eq.~(\ref{sfexp}) grows. In fact, it can be shown both
numerically\cite{brus} and analytically\cite{sxap}
 that the ratio of the NL contribution to the splitting function to
the leading order 
grows linearly as a function of $\Xi$, and in fact  
overwhelms the leading order for any reasonable value of $x$ and 
$Q^2$ (fig.~7). The problem is in 
fact worse in ``physical'' schemes 
where the ratio grows as $\Xi^{3/2}$; the same happens
in the so-called $Q_0$ scheme\cite{ciaqz} (designed to reduce perturbative
corrections in the quark sector), as well as in the
 momentum conserving scheme\cite{mom}, though the slope of the growth
is somewhat smaller in the latter case.

The growth of the  NLO ``correction'' with respect to the leading
order means  that
the NLO $P^{(1)}$ eq.~(\ref{sfexp}), 
though formally subleading, 
is actually enhanced at large  $\Xi$  with respect to $P^{(0)}$: 
even though it is suppressed by an extra power of $\alpha_s$, at
sufficiently small $x$ it always
takes over due to the growth of $\Xi$ with $\ln {1\over x}$ at fixed
$\alpha_s$.
Hence,  it is the perturbative expansion of the
leading $\ln {1\over x}$ contributions to the Altarelli-Parisi
equations 
defined by  eq.~(\ref{sfexp}) which breaks down.

On the one hand, this is good news, in that it explains why the inclusion
of the leading $\ln{1\over x}$ summation $P^{(0)\,ij}$ spoils
the agreement with the data: these terms are actually not leading in
the HERA region, and thus there is no reason why including them while
not including more important contributions should improve the
phenomenology. On the other hand, however, the fact that
standard NLO works at all --- not only at HERA, but also in the SLAC
or NMC region (see Fig.~7) --- is now puzzling:
there must be very large cancellations in the anomalous dimension
eq.~(\ref{asexp}), or, otherwise stated, 
the expansion eq.~(\ref{sfexp}) is pathological. 

It is thus necessary to reorganize the expansion. Now, it can be
shown\cite{sxap}  
that the rise of the NL correction is entirely generic --- it
could have been predicted without waiting the outcome of the
Fadin-Lipatov calculation: the ratio of the N$^k$LO to the LO 
splitting functions
eq.~(\ref{sfexp}) behaves as ${P^{(k)}\over P^{(0)}}\sim
[\alpha_s\ln({1\over x})]^{2k-1}$. This behavior is related to the
fact that generically
the anomalous dimensions $\gamma^{(k)}$ eq.~(\ref{asexp}) 
has a (2k-1)---th
order pole at $ {C \alpha_s\over N}=1$. The slope of the  $k$-th order
rise is determined by the residue of the corresponding pole, which can in
turn be simply expressed in terms of the $k$--th order corrections to
high-energy asymptotics. 

It is possible to eliminate this singularity in the anomalous
dimension, and thus the instability of the perturbative expansion, by
suitably choosing the factorization scale  order by order in
$\alpha_s$. This is
however a fine-tuning: even small variations 
of the factorization scale around
the optimal scale bring back  immediately the instability (see Fig.~8)
A better
option consists of reabsorbing into the leading order $P^{(0)}$ all
contributions which lead to the small $x$ growth of
the formally subleading $P^{(i)}$. It turns out that this can be done
in a completely unambiguous way by subtracting terms related to the
residues of the poles in the anomalous dimensions. 
\begin{figure}
\centering\mbox{
    \epsfig{file=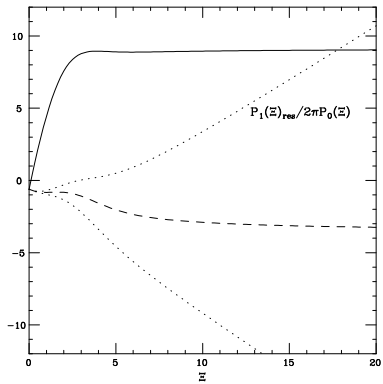,width=6.cm}}
 \caption[]{The ratio of NLO to LO splitting functions eq.~(11) in the
    DIS scheme after
    resummation (solid), with fine-tuned scale (dashed) and scales a
    factor of two either side of the fine-tuned one (dotted).
(from Ref.\cite{sxap})}
\end{figure}

The overall
ambiguity introduced by the procedure is contained in a single
parameter $\lambda$, 
which could only be determined by
resumming the splitting function to all orders. This parameter governs
the all-order asymptotic behavior of the splitting
function $P(x,Q^2)$ eq.~(\ref{sfexp}): $P(x,Q^2)\sim
(\ln{1\over x})^{3/2}
x^\lambda$. The ensuing splitting functions can be viewed as a resummation of the
original series: the leading order resummed $P^{(0)}$ depends on the
all-order resummed
parameter $\lambda$, but the ratios of all higher order $P^{(i)}$ to $P^{(0)}$
is fixed, and tend asymptotically to a constant as $x\to 0$ (Fig.~8).

In an Altarelli-Parisi framework this does not entail
loss of predictivity, since the $Q^2$-independent part of the 
$x$ dependence of the
structure function cannot be determined anyway. 
However, it does
show that even assuming that the asymptotic small $x$ behavior of structure
functions is determined by  perturbative QCD asymptotics, it cannot
be determined in any finite order computation and would require an
all-order resummation of the small-$x$ expansion.

\subsection{Energy evolution}
The  resummation of the 
the leading $\ln{1\over x}$ contributions to structure
functions discussed in the previous section is not the end of the
story, however. So far, we have discussed leading $\ln{1\over x}$
contributions
which are generated when 
solving evolution equations in $Q^2$, i.e. the 
Altarelli-Parisi equations. 
This, however, is an indirect manifestation of a more fundamental
phenomenon, namely, the fact that the high-energy behavior of the
gluon-gluon scattering cross section in perturbative QCD
is dominated by energy logs. 

In DIS energy logs  appear as $\ln W^2\sim \ln
{1\over x}$, and  can be summed  by means of an equation for evolution
in $1\over x$. This is
accomplished by the BFKL equation,\cite{vddrev} which in the fully inclusive case
can be written in the same form as the usual QCD evolution
equation, but
with the roles of the two kinematic variables $x$ and $Q^2$
interchanged. To this purpose, it is necessary to 
take a Mellin transform with respect to $Q^2$
and define kernels $\chi(M)$
as functions of the associate
variable $M$, which can then be used to evolve in $\xi\equiv\ln{1\over
x}$:
\bea
\label{sxee}
&&{d\over d\xi} G(x, M)=\alpha_s \chi(M) G(x,M)\nonumber \\
&&\quad G(x,M)\equiv \int_0^\infty {d Q^2\over Q^2} 
\left({Q^2\over\Lambda^2}\right)^{-M} G(x,Q^2)\\
&&\quad \chi(M)=\chi^{(0)}(M)+\alpha_s \chi^{(1)}+\dots ,\nonumber
\eea
where  $G(x,Q^2)$ is the leading eigenvector of the anomalous
dimension matrix eq.~(\ref{lsexp}) 
(i.e. the one associated to the only nonvanishing leading--order
eigenvalue), which at leading order is just
$G(x,Q^2)=xg(x,Q^2)$ 
with $g(x,Q^2)$ the gluon distribution.\footnote{Eq.~(\ref{sxee})
applies at fixed coupling $\alpha_s$.
Beyond leading
order the running of the coupling $\alpha_s$
with $Q^2$ must also be included;
this introduces some technical complications in the form of
eq.~(\ref{sxee}) and its solution which are however inessential for
our discussion.}
Since $G(x,Q^2)$ behaves as a constant at large $Q^2$ 
(Bjorken scaling)
and falls linearly as $Q^2\to0$ (for kinematic reasons\cite{books}),
the (leading twist) physical region in $M$ space is
$0<M<1$.  At leading order,
$\chi^{(0)}(M)=2 \gamma -\psi(M)-\psi(1-M)$ (where 
$\psi(M)$ is the
digamma function) ---
the celebrated BFKL kernel  --- and
the recent FL calculation\cite{fl} in the 
inclusive case amounts to a
determination of the next-to-leading correction $\chi^{(1)}(M)$.
\begin{figure}
\centering\mbox{
    \epsfig{file=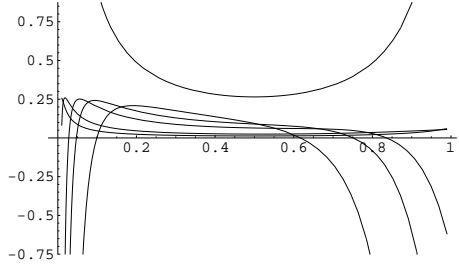,width=6.cm}}
 \caption[]{The kernel $\chi(M)$ eq.~(\ref{sxee})
plotted as a function of
    $M$. The curves from top to
    bottom in the middle are the LO result with $\alpha_s=0.1$
and the NLO result with $\alpha_s=0.1$, 0.05, 0.03, 0.01, 0.005.
(adapted from Ref.\cite{brus})}
\end{figure}

The qualitative behavior of the kernel is
changed dramatically by the NL correction 
(see Fig.~9). To understand the effect of this change, solve the
evolution equation 
eq.~(\ref{sxee}) and construct $G(x,Q^2)$ by inverting the Mellin
transform:
\be
G(x,Q^2)=\int_{-i\infty}^{i\infty}\!{dM\over2\pi
i}\,\left({Q^2\over\Lambda^2}\right)^M e^{\xi \chi(M)} G_0(M).
\label{sxsol}
\ee
The asymptotic behavior of the integral
at small $x$, i.e. large $\xi$ is determined 
by saddle point.
Since the integration path runs along the imaginary axis, 
the saddle is
a minimum of $\chi(M)$ along the real axis, which at leading order
is located at 
$M={1\over2}$, where $\chi^{(0)}(1/2)=4\ln 2$. But at NLO
there is no real minimum, and the real saddle is thus replaced by two
complex saddles (which must occur in complex conjugate
pairs by definition of the kernel as the Mellin transform
of a real splitting function).\cite{brus,ross} 
Hence the NLO, rather than being a small correction to the LO, 
changes the asymptotic behavior, and in fact leads to an oscillatory,
unphysical behavior at large $\xi$. In particular, the
value of $\chi(1/2)$ has  no special meaning if
$\chi$ is computed at NLO. Of course, at small enough $\alpha_s$
i.e. high enough scales, the
leading
order behavior is recovered: however, this only happens (Fig.~9) at
values of $\alpha_s\lsim 0.01$ 
which correspond to a scale above the Planck
mass.\cite{brus}

The original calculation on which these results are based,\cite{fl} 
which is the
result of an effort of years, has been in large part checked independently: 
most of the individual real and virtual amplitudes have been
recomputed with different techniques;\cite{vdd} and the elaborate
computation that takes from the individual amplitudes to the kernel 
has also been checked.\cite{ciaf} 
It thus appears unlikely
that the result\cite{fl} is erroneous.
The saddle point estimate of the asymptotic behavior has also been
checked by more accurate calculations, which have confirmed the
unphysical oscillatory behavior of the 
cross--section.\cite{lev,bart} 

The problem posed by this unphysical behavior is separate from that
discussed in Sect.~4.1. Indeed, 
there we saw
that the naive small $x$ expansion of the anomalous dimension 
must be reorganized in such a way that the terms which contribute to
the asymptotic small $x$ behavior to all orders in the 
expansion eq.~(\ref{lsexp}) 
are resummed into a parameter $\lambda$, and
included into  the leading order of a resummed
expansion.  Now, we are studying how the all-order asymptotic behavior
can be derived from $\chi$, and
we see that if $\chi$ is computed to NL order, the associate
asymptotic behavior is unphysical. It therefore looks
like a further resummation of  $\chi$ is required.

Several partial resummations of formally
subleading corrections to $\chi$  have thus been suggested. 
First, one can attempt to resum logs
of $Q^2$ related to the running of the coupling\cite{muel}, though
this actually seems to make things worse.\cite{bart}
Also, one may try to optimize the scale choice by the BLM 
method\cite{bfklp}:
this goes in the right direction, but does not completely remove the
instability. 
A further option consists of noting
that the definition of the leading energy logs which are being
resummed requires
a choice of scale: for instance, in DIS at small $x$, $s=W^2\approx {Q^2
\over x}$, and the resummation of $\ln{1\over x}$ may be more
generally viewed as a resummation of
$\ln {s\over s_0}$ with
$s_0=Q^2$. Different choices of
$s_0$ then lead to different subleading corrections, and a resummation
of these ``scale-dependent'' corrections may change the properties of
the evolution equation eq.~(\ref{sxee}).\cite{salam}
A specific resummation of such corrections has been advocated on the
grounds that it removes some undesirable singularities of $\chi(M)$ at
$M=0$, thus  leading to an improved asymptotic
behavior\cite{salam,ciares}. This does not, however, lead
to an unambiguous prescription.
 Use of an
evolution equation based on angular ordering 
seems to further increase
the theoretical ambiguities.\cite{ccfm}

A perhaps more fundamental possibility~\cite{lipres,schm}
is to  re-examine the assumptions on the high-energy asymptotics that
go 
into the NL determination of $\chi$. In particular, by introducing
physically motivated kinematic cuts in the course of the determination
of $\chi(M)$ 
one can reshuffle contributions between the NL $\chi^{(1)}$ and the LO
kernel $\chi^{(0)}$ while only changing the (unknown) NNL
terms. Again, the shortcoming of the approach is that it introduces an
ambiguity, in the form of a
dependence on a ``cutoff'' parameter. However, 
the ensuing NL result is then well-behaved in a reasonable range of
values of this cutoff. The ambiguities can be somewhat 
reduced\cite{forsh}  by again resumming ``scale dependent''
corrections along the lines of Ref.\cite{salam}

In conclusion, the  ambiguities  encountered when trying to
construct an energy evolution equation appear to be
related to the choice of  the appropriate large
scale. Indeed, on the one hand the large log which is treated as
leading and
summed by the
evolution equations  is $\xi$, associate to the large scale $W^2$, but
on the other hand the hard scale which guarantees the validity of
perturbative factorization and the smallness of $\alpha_s$ is
$Q^2$. The need to reconcile these two large scales appears
to be at the
origin of the difficulties of perturbation theory at small $x$. A
radical suggestion, based on the idea that $W^2$ should also be used
as a factorization scale, was discussed in ref.\cite{afp};
a proof of the corresponding factorization
theorems would however be required in order for
this approach to be viable beyond leading order.
We must conclude that no reliable theory of small $x$ evolution is
yet available. The success of simple NLO perturbation theory and the
small $x$ approximation to it still lacks a completely 
fully satisfactory explanation.

\section{ The embarrassing success of QCD }
The application of perturbative QCD to inclusive deep-inelastic
scattering is embarrassingly successful. In many ways, QCD is enjoying
the same sort of success as the electroweak sector of the standard
model. Whether the lack of evidence for nonstandard effects should be
considered a triumph or a disappointment is of course matter of
opinion. However, in many instance things are now too easy: we do not
see deviations from the simple leading log behavior even in kinematic
regions where we
might have expected 
them, in particular at very large and very small $x$.
This reveals that our understanding of perturbative QCD and its
limitations is still not satisfactory from a theoretical viewpoint.

\section*{Acknowledgments}

I thank M.~Greco for giving me the opportunity to
present this review in the beautiful setting of La Thuile, 
 G.~Altarelli for several discussions and R.~D.~Ball
for a critical reading of the manuscript.


\begin{thebibliography}{99}
\bibitem{books} R.K.~Ellis, W.J.~Stirling and B.~Webber, {\em QCD and
Collider Physics} (Cambridge U.P., Cambridge, U.K.,
1996); R.~G.~Roberts,
{\em The structure of the proton} (Cambridge U.P., Cambridge, 
U.K., 1990)
\bibitem{guido} G.~Altarelli, {\em Phys. Rep.} {\bf 81}, 1 (1982)
\bibitem{bcdms}  BCDMS Coll., A.~Benvenuti et al., {\em
Phys. Lett.} {\bf B223}, 485 (1989)
\bibitem{nmc} NMC Coll., M.~Arneodo et al., {\em Phys. Lett.}
{\bf B364}, 107 (1995)
\bibitem{slac} L.~W.~Whitlow et al., {\em Phys. Lett.} {\bf B282}, 475 (1992) 
\bibitem{h1van} H1 Coll., presented by A.~De~Roeck at ``Nucleon '99'',
Frascati, June 1999.
\bibitem{tul} See e.g. SMC Coll., B.~Adeva et al., Phys. Rev. {\bf D58}, 112001
(1998); T.~\c Cuhadar, Ph. D. Thesis, Amsterdam Free University, 1998
\bibitem{mods} M.~Rueter, {\tt hep-ph 9807448};
A.~Donnachie and 
P.~V.~Landshoff, {\em Phys.Lett.} {\bf B437}, 408 (1998);
P.~Desgrolard, A. Lengyel and E.~Martynov, {\tt hep-ph/9811380}
\bibitem{haidt} W. Buchm\"uller and  D. Haidt, {\tt hep-ph/9605428};
 D. Haidt, talk at the DIS99 meeting
\bibitem{abfr} G.~Altarelli et al., {\em Nucl. Phys.} {\bf B496}, 337
(1997); {\em Acta Phys. Pol.} {\bf B29} 1145 (1998)
\bibitem{test}  R.~D.~Ball and S.~Forte, {\em
Phys. Lett.}  {\bf B336}, 77 (1994) 
\bibitem{diff} For reviews see E.~Predazzi, {\tt hep-ph/9809454};
A.~Hebecker, {\tt hep-ph/9905226};  W.~Buchm\"uller, {\tt hep-ph/9906550}
\bibitem{frac} L.~Trentadue and G.~Veneziano, {\em Phys. Lett.} {\bf
B323}, 201 (1993)
\bibitem{deflor} D.~De Florian and  R.~Sassot, {\tt hep-ph/9808300}
\bibitem{pdb}  Particle Data Group, C.~Caso et al.,
{\em Eur. Phys. J.} {\bf C3}, 1
(1998) 
\bibitem{pdbold}  
Particle Data Group, R.~M.~Barnett et al, 
{\em Phys. Rev.} {\bf D54}, 1 (1996) 
\bibitem{asex} H1 Coll., T.~Ahmed et al, {\em Phys. Lett.} 
{\bf B346}, 415 (1995); 
ZEUS Coll., M.~Derrick et al, {\em Phys. Lett.},
{\bf B363}, 201 (1995); H1 Coll., subm. to ICHEP98 (Vancouver, July
1998), Abstract 528
\bibitem{vm} M.~Virchaux and A.~Milsztajn, {\em Phys. Lett.} {\bf
B274}, 221 (1992)
\bibitem{nmcas} NMC Coll.,
M.~Arneodo et al., {\em Phys. Lett.} {\bf B309} 222
(1993)
\bibitem{ccfr} CCFR Coll.,
W.~G.~Seligman et al., {\em Phys. Rev. Lett.} {\bf 79},
1213 (1997)
\bibitem{ccfrold} CCFR Coll.,
P.~Z.~Quintas et al., {\em Phys. Rev. Lett.} {\bf
71}, 1307 (1993).
\bibitem{nnlo} S.~A.~Larin, T.~van~Ritbergen and J.~A.~M.~Vermaseren,
{\em Nucl. Phys.} {\bf B427}, 41 (1994);
{\em Nucl.Phys.} {\bf B492}, 338 (1997)    
\bibitem{kata} A.~L.~Kataev et al., {\em Phys. Lett} {\bf B417}, 374
(1998) 
\bibitem{katanew}
A.~L.~Kataev, G.~Parente and
A.~V.~Sidorov , {\tt hep-ph/9905310}. 
\bibitem{ynd} J.~Santiago and F.~J.~Yndurain, {\tt hep-ph/9904344} 
\bibitem{ale} S.~Alekhin, {\em Phys.Rev.} {\bf D59}, 114016 (1999) 
\bibitem{alkat} S.~I.~Alekhin and A.~L.~Kataev 
{\em Phys. Lett.} {\bf B452} 402 (1999) 
\bibitem{polrev} See e.g. S.~Forte, {\tt hep-ph/9610238};
R.~D.~Ball, {\tt hep-ph/9812383}
\bibitem{polhera} A. De Roeck et al., {\em Eur.
Phys. J.} {\bf C6},121 (1999)   
\bibitem{gerr} J.~Huston et al., {\em Phys. Rev.} {\bf D58},
114034 (1998)    
\bibitem{bfr} R.~D.~Ball, S.~Forte and G.~Ridolfi, {\em Phys. Lett.}
{\bf B378}, 255 (1996)
\bibitem{nnnlo} S.~A.~Larin and J.~A.~M.~Vermaseren, {\em Phys. Lett.}
{\bf B259}, 345 (1991)
\bibitem{elkar} J.~Ellis et al., {\em Phys. Lett.} {\bf B366}, 268 
(1996)
\bibitem{ccfrgls} CCFR Coll., J.~H.~Kim, D.~A.~Harris et al.,
{\em  Phys. Rev. Lett.} {\bf 81}, 3595 (1998)
\bibitem{dgptwz} A.~de~R\`ujula et al., {\em Phys. Rev.} {\bf D10},
1649 (1974)
\bibitem{oldnmc} NMC Coll., M.~Arneodo, {\em Phys. Lett.}
{\bf B295}, 159 (1992) 
\bibitem{das} R.~D.~Ball and S.~Forte, {\em Phys. Lett.} {\bf B335},
77 (1994) 
\bibitem{zako} R.~D.~Ball and S.~Forte, {\it Acta Phys. Pol. } {\bf 
26}, 2079 (1995)
\bibitem{lech} L.~Mankiewicz, A.~Saalfeld and T.~Weigl,
{\em  Phys. Lett.} {\bf B393}, 175 (1997)
\bibitem{alphaold} R.~D.~Ball and S.~Forte, {\em Phys. Lett.} {\bf
B358}, 365 (1995)
\bibitem{alphanew} R.~D.~Ball and S.~Forte, {\tt hep-ph/9607289}
\bibitem{mrstht} A.~Martin et al., {\tt hep-ph/9808371} 
\bibitem{cteq}  CTEQ Coll., H.~L.~Lai et al., {\em Phys.
Rev.} {\bf D55}, 1280 (1997),; W.~K.~Tung, {\tt hep-ph/9608293} 
\bibitem{mrst}   A.~Martin et al., {\em Eur. Phys. J.}
{\bf C4}, 463 (1998)
\bibitem{ucc} A.~Piccione and G.~Ridolfi, {\em Nucl. Phys.}
{\bf B513}, 301 (1998)
\bibitem{llx} D.~Amati et al., {\em Nucl. Phys.} {\bf B173}, 329
(1980) 
\bibitem{nllx} S.~Catani and L.~Trentadue, {\em Nucl. Phys.} 
{\bf B327}, 323 (1989); 
G.~Sterman, {\em Phys. Lett.} {\bf B179}, 281 (1986)  
\bibitem{lxope} R.~Akhoury, M.~G.~Sotiropoulos and G.~Sterman,
{\em Phys. Rev. Lett.}, {\bf 81}, 3819 (1998)
\bibitem{benrev} M.~Beneke, {\tt hep-ph/9807443} and ref. therein
\bibitem{sfren} E.~Stein et al, {\em Phys. Lett.} {\bf B376}, 177
(1996); M.~Dasgupta and B.~R.~Webber, {\em B382}, 273 (1996);
M.~Meyer-Herrmann et al., {\em Phys. Lett.} {\bf B383}, 463 (1996)
\bibitem{rensing} E.~Stein et al., {\tt hep-ph/9803342}
\bibitem{ugo} U.~Aglietti, {\em Nucl. Phys.} {\bf B451}, 605 (1995) 
\bibitem{bodek} U.~K.~Yang and A.~Bodek, {\tt hep-ph/9809480};
U.~K.~Yang, 
talk at ICHEP98 (Vancouver, July
1998), {\it unpublished}
\bibitem{mrs} A.~D.~Martin, R.~G.~Roberts and J.~Stirling, {\em
Phys. Lett.}  {\bf B387}, 419 (1996)
\bibitem{zij} E.~B.~Zijlstra and W.~L.~van~Neerven, {\em
Nucl.Phys.} {\bf B383}, 525 (1992) 
\bibitem{liuti} S.~Liuti, {\tt hep-ph/9809248} 
\bibitem{uvdom} M.~Beneke, V.~M.~Braun and L.~Magnea, {\em
Nucl. Phys.} {\bf B497}, 297 (1997) 
\bibitem{htmod} X.~Guo and J.~Qiu, {\tt hep-ph/9810548}
\bibitem{simula} G.~Ricco, S.~Simula  
and M.~Battaglieri {\tt hep-ph/9901360}
\bibitem{lgad} T.~Jaroszewicz, {\em Phys. Lett.} {\bf
B116}, 291 (1982);
S.~Catani, F.~Fiorani and G.~Marchesini, {\em Phys. Lett.} {\bf B336},
18 (1990); S.~Catani et al.,  {\em Nucl. Phys.} {\bf B361}, 645 (1991) 
\bibitem{vddrev} For a review, see  
V.~del~Duca, {\tt hep-ph/9503226} and ref. therein
\bibitem{levwr} See e.g. E.~Levin, {\tt hep-ph/9503399} 
\bibitem{afp} R.~D.~Ball and S.~Forte, {\em Phys. Lett.} {\bf
B405}, 317 (1997)
\bibitem{summ} R.~D.~Ball and S.~Forte, {\em Phys. Lett.} {\bf
B351}, 313 (1995)
\bibitem{ehw}
R.K.~Ellis, F.~Hautmann and B.~R.~Webber, {\em Phys. Lett.}{\bf B348},
{582} (1995)
\bibitem{nlqad} S.~Catani and F.~Hautmann, {\em Phys. Lett.} {\bf
B315}, 157 (1993); {\em Nucl. Phys.} {\bf B427}, 475 (1994)  
\bibitem{roma} R.~D.~Ball and S.~Forte, {\tt hep-ph/9607291}
\bibitem{phys} S.~Catani,  {\em Z. Phys.} {\bf C75}, 665 (1997)
\bibitem{fl} V.~S.~Fadin and L.~N.~Lipatov, {\em Phys. Lett.}
{\bf B429}, 127 (1998) and ref. therein
\bibitem{brus} R.~D.~Ball  and S.~Forte, {\tt hep-ph/9805315}
\bibitem{blum}  J. Bl\"umlein et al., {\tt hep-ph/9806368}
\bibitem{sxap} R.~D.~Ball  and S.~Forte, {\tt hep-ph/9906222}
\bibitem{ciaqz} M.~Ciafaloni, {\em Phys. Lett.} {\bf B356}, {74} (1995)
\bibitem{mom} R.~D.~Ball and S.~Forte, {\em Phys. Lett.} {\bf
B359}, 362 (1995)
\bibitem{ross} D.~A.~Ross, {\em Phys. Lett.} {\bf B431},
161 (1998) 
\bibitem{vdd} V.~del~Duca, {\em Phys. Rev.} 
{\bf D54}, 989 (1996); {\em Phys. Rev.} 
{\bf D54}, 4474 (1996); V.~del~Duca and C.~R.~Schmidt, 
{\tt hep-ph/9810215};
Z.~Bern, V.~del~Duca and C.~R.~Schmidt, {\tt
hep-ph/9810409} 
\bibitem{ciaf} G.~Camici and M.~Ciafaloni, 
{\em Phys. Lett.} {\bf B412}, 396 (1997); {\em  Nucl. Phys.}
{\bf B496}, 305 (1997);  {\em Phys. Lett.} {\bf B430}, 349
(1998)
\bibitem{lev} E. Levin, {\tt hep-ph/9806228} 
\bibitem{bart} N.~Armesto, J.~Bartels and M.~A.~Braun, {\tt
hep-ph/9808340}
\bibitem{muel} Y.~V.~Kovchegov and A.~H.~Mueller, {\tt 
hep-ph/9805208}
\bibitem{bfklp} S.~J.~Brodsky et al, {\tt hep-ph/9901229}
\bibitem{salam} G.~P.~Salam, {\em JHEP} {\bf 07}, 19 (1998) 
\bibitem{ciares} M.~Ciafaloni and D.~Colferai, {\tt hep-ph/9812366} 
\bibitem{ccfm} G.~Bottazzi et al, {\tt hep-ph/9810546}
\bibitem{lipres} L.~N.~Lipatov, talk at the FNAL small $x$ workshop (1998)
\bibitem{schm} C.R.~Schmidt, {\tt hep-ph/9901397}
\bibitem{forsh} J.~R.~Forshaw, D.A.~Ross and A.~Sabio~Vera, {\tt
hep-ph/9903390}
\end{thebibliography}
\end{document}